# FROM DIGITAL COMPUTERS TO QUANTUM COMPUTERS BASED ON BIOLOGICAL PARADIGMS AND PROGRESS IN PARTICLE PHYSICS

P.N. Borza, Transylvania University (Brasov);  L.F. Pau, CBS (Copenhagen)  and Erasmus University (Rotterdam)  lpau@nypost.dk

## Abstract

While several paths have emerged recently in microelectronics and computing as follow-on's to Turing architectures  implemented using essentially silicon circuits, very little  "Beyond Moore" research has considered, first biological processes instead of sequential instructions, and, next the implementation of these processes exploiting particle physics interactions. This combination allows e.g. native spatio-temporal integration and correlation, but also powerful interference filtering / gating / splitting and more. These biological functions, their realization by quantum and charge carrier interactions, allow proposing a novel computing architecture, with interfaces, information storage, and programmability. The paper presents the underlying biological processes, the particle physics phenomena which are exploited, and the proposed architecture, as well as an algebraic design formalism.

## Introduction

The present paper intends to suggest the necessity to evolve the basic paradigm for digital processing systems, in the context of ever more apparent performance limitations and of some of the discoveries made in biology. The prime thrust is to hinge upon progress in particle physics, to capitalize on ways quantum processing allows to replicate biological processes with resulting computing performance benefits. Whereas a few authors have envisioned the potential for quantum physics in computing [1-2], they did not link it to progress in biology and only to a limited extent to recent discoveries in particle physics. As this scope is quite wide, the focus here will be to address some of the architectural elements, their realization, and to relate them to functional biological processes.

We cover the following aspects:

i)   an overview of the basic biological and physiological phenomena exploited in the suggested proposal;
ii)  survey advances in research on sub-particles with an emphasis on the implementation of new specific functionalities inspired by i);
iii) describe the resulting architectural building blocks;
vi)  analyze some of the benefits to be expected;
vii) discuss open research and technological issues.

The conclusions summarize the main ideas in the paper, and also suggest a path towards a theoretical development of a generic architectural simulation model based on colored algebras.

The intent with this paper is not to discuss in detail each constituent of the architecture, or the physics and biology which are exploited, but to offer an integrated vision of their interplay, and how they can be assembled to enable a novel type of information processing systems.

# I. Survey

This brief survey contrasts conventional digital computer architectures and technologies, with examples of adaptative biological processes, and shows how quantum physics in new materials have led to very relevant new processes (discussed in more detail in Section V).

From an historical point of view the last past 80 years were mainly influenced by the computer development based on fundamental research done by Turing [3]. The bases for Turing machines were binary logic and several very simple structures, that combined realized two main functions: storage of the information in a binary form (1 digit), and the implementation of elementary logical and arithmetical operations. The finite state machine is the base of modern computers [4], but Turing also introduced "instruction based" systems, enabling the programmability of such systems. So, from practical point of view, the electronic computers were born as implementation of ideas issued in 1938 [3]. The instruction sequence decides upon functionality [4-5]. The physical implementation was carried out as a result of long technological development efforts, initially using electronic tubes and later electric switching phenomena in semiconductor materials. These phenomena produce a significant amount of heat especially during transient regimes. With clock frequency and functional density in silicon both increasing, the "thermal wall" combined with input-output bandwidth limitations, represent nowadays major obstacles for significant performance gains [6-7].

Looking in contrast at biological systems, it has become obvious that the gating process that happens in neurons at synapses level can also handle sophisticated information, relying on spatial-temporal integration and correlation. The threshold potential function ensures the filtering of noise that could appear on the propagation path of the nervous signal. A single remark is important here, i.e. that biological systems have already realized the fundamental and encapsulated link between the status of neurons and time. The topology of nervous systems, illustrated by the huge variety of ganglions inserted on signal pathways, represents complex organic multiplexors and de-multiplexors. In neural pathways, the biochemical and electrical signals' walk and mutual adaptation of the sensitivity thresholds play the essential role in preserving the stability, reliability and "filtering" capacity of the neural processing [8].

Example 1: Vision analysis: This co-existence of coordinated heterogeneous biological processes is relevant, for example, in the case of the visual analyzer where the "lateral inhibition" phenomena is also present, consisting in a cortical "reflection" process between different cortical zones [9] and the afferent visual ways, illustrating the necessity of processing in a reverberant way of spatial distributed bulk signals.

Example 2: Allostatic load: This "wear and tear process" on the body due e.g. to stress, was first studied by Sterling and Eyer [8-9].It achieves stability, or homeostasis, through physiological and behavioral change. It alters HPA axis hormones, the nervous system, cytokines, and is generally adaptive in the short term. In these cases, the parameters vary and the variation anticipates the demands. This means in fact that the specific, dynamically modified and self-updated parameters, which are managed by a biological system, play a fundamental role in achieving a remarkable reliability, stability and variety in systems' adaptation and response. The prediction processes require that each sensor will exhibit an adaptation in the optimal input range [10].

Example 3: Cortical reflections: The cortical area is responsible for very complex phenomenon related to thinking, feeling, rewarding/pain, and fusion that happen at cortex level [11].The most interesting aspect of the cortex is related to the "reverberant" loops which can be used as simple clock generators, or for spatio-temporal filtering of the charge carriers beams, or to label information items by cortical zones for data structuring. The maps of these zones have already been identified [12]. Selective "reflective" functional blocks mimicking even limited properties of the cortex, can carry out essential information handling and processing, and their realization can use selective Moiré filtering (Figure 5).

Example 4: Heart conduction system: The cardiac conduction system (CCS) exhibits a typical "hard coded" control function, where the propagation of muscular depolarization pulses is along path flow and the propagation delays are the control variables for the blood pump function of the heart [13]. Also, the CCS is a realization of a fault-tolerant hierarchical control architecture. The combination of the different control signal paths and the distribution of heart muscle fibers, functioning as effectors, generate a highly adaptive system regulator.

Finally, recent discoveries in fundamental particle physics, such as quantum effects in carbon allotropic phases, have led to proofing logical functions implemented in graphene and carbon nanotubes (CNT) [14-17]. This generates the premises for changing the much known paradigms related to the conduction of charge carriers [18-20]. As an example, graphene has been shown to exhibit altogether thermal conductivity with the $k$ values of 5,000W/mK, Young module of 1TPa [21], and also very high electro conductivity, thus allowing the easy implementation of digital and memory functions [23]. As another example, recent studies revealed that the deflection of charge carriers appear as the result of the existence of different kinds of lattice places on their pathways [24]; the lattice potential is able to deflect the charge carrier beams. Such lattices can be implemented in silicon, but also on graphene structures using effects such as Moiré patterning. This can be implemented using the Quantum Hall Effect (QHE), [19, 25] which deflects the charge carriers from their straight path; two 2D-layers of the phase form a network that acts as Moiré filter and deflect the charge carriers flow.

Over the past approx. 10 years, research has been devoted to "unconventional computing paradigms", which is the terminology chosen by the European Research consortium for informatics and mathematics to cover such approaches as molecular and cellular computing, quantum computing, neural network processing [26] .Likewise, IBM with its project Synapse devoted, and others, have realized dedicated chips in CMOS processes. However, in contrast to the present paper, all of them still only consider implementation on silicon, and very few address architectural, programming and integration aspects. Already now, some silicon photonic building blocks have been implemented, which progressively should get migrated to quantum flows in graphene or AIN photonic building blocks: ring resonators [27], whispering gallery mode resonators [28], directional couplers [29], grating couplers [30], slot waveguides [31], self-adjusting Mach-Zehnder interferometers [32], and photonic crystals [33]. They enable limited versions of optical computing. In the MOQUASIMS project [34], has been realized a quantum system capable of capturing flying photons and hold them in stationary atomic excitations akin optical memory storage. However, the manufacturability of such components poses formidable challenges, which imply that the proposed architecture would in contrast relatively benefit from the use of high volume graphene or CNT materials with more stable manufacturing processes.

The above survey justifies the approach taken in this paper is to exploit the potential of quantum effects in non-optronic materials, for the implementation of selected biological processes, with a controllability and programmability inherited from Turing machines.

## II. Research question

The aim of the paper is to outline a potential architecture for the next information processing system generation. Relying on quantum information, and biologically inspired functional blocks, with electronic interactions in magnetic materials, it is conceived using the conduction, deflection, filtering and association phenomena in the operation of nanostructures mimicking biological processes triggered by particles interactions.

# III. A basic quantum biological architecture

In this Section we first briefly introduce the hypothesis underlying the envisaged quantum-biological digital processing architecture, as illustrated in Figure 1, and further detailed in Section VI.:

<u>Hypothesis 1</u>: the processing is done by information-carrying quantum particles interacting along propagation pathways. These massless quantum energy carriers may co-exist with energized sub-particles able, based on their energy, to propagate along pathways. Deflection or collection elements, specific to the carrier types, may be implemented along the pathways. The dual particle types also enable different delays.

<u>Hypothesis 2</u>: the transformation of the quantum particles is realized by functions which can mimic biological or physiological processes; examples presented later in this paper include differential/integral operators, deflection lattices, and reflective processes;

<u>Hypothesis 3</u>: the signalling amongst the pathways, or in sub-lattices, is relying on magnetic-electrical interactions, as recently discovered [35];

<u>Hypothesis 4:</u> the spatio-temporal filtering of charge carriers exploits the Moiré processing, or structured resonators;

<u>Hypothesis 5</u>: information storage is realized by resonant quantum structures, such as quantum wells or quantum dots, capable of producing single photons on demand at high rates [36];

<u>Hypothesis 6</u>: input/input is via asynchronous quantum charge generation at very high data rates, or synchronous read-out by conventional means at lower data rates;

<u>Hypothesis 7</u>: the programming is realized in a conventional digital CPU with photonic outputs activating the control/ signalling by particles.

Presented as summarized above, the envisaged architecture involves a conventional command node for legacy and programmability reasons, but coupled to a novel processing system. In this processing system, quantum charge accumulation, distribution and decay are executed via by the functional blocks inspired from biology. The quantum pathways are changed by the functional blocks. The control path gives the software driven sequence of functions and parameters to be executed. The signalling path, separate from the control path, activates/ deactivates the set of required functional blocks and pathways.

In physics terms, the quantum pathways involve three different zones:

-charge carrier confinement areas, with gates emitting charge carriers (controlled by the signalling) and where input synchronization is carried out;

-the actual quantum pathways, where the charge carriers will propagate; interference may occur at selected locations via charge collectors;

-charge carrier collection areas, where outputs are generated, switched and synchronized

The proposed architecture mimics also biological systems. Taking the Example 4 of Section II, regarding the cardiac conduction system, the quantum pathway is analogous from the pathway from the sino-atrial node, to the atrio-ventricular path segment, to the Hiss conduction channel, and finally to Purkinje terminations. With these propagation channels, the heart becomes a muscular pump with different delays in the different pathways, when depolarization control impulses reach the heart muscular cells

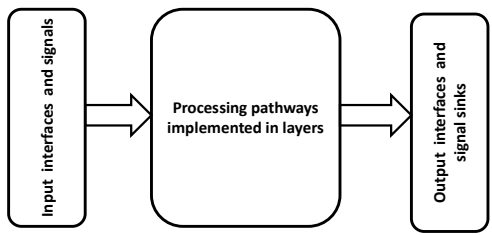

Figure 1: Basic proposed quantum biological architecture

## IV. Physics phenomena exploited by the proposed architecture

At the core of the proposed architecture, lie the functional blocks. The properties thereof in terms of quantum pathways transformations rely first on a geometrical structure, and next on the selection / injection of selected defect categories. The phenomena taking place are effects such as ballistic conduction, the confinement of the charge carriers (electrons and positrons) in accordance with fermions composition rules. These phenomena are induced by structural defect inside the 1D, 2D and multi-D allotropic structures, and/or by group effects. These are the result of building up, at nano-scale of successive and different 2D layers; in this way can be designed new variants and also very complex processing structures. The 2D layers can be based on graphene, boron-nitride, or other ferromagnetic nano-structured materials, which can be inserted inside regular structures [37]. The succession of these layers will permit to deflect the charge carriers' spots, and add sensing local loops.

For the control and signalling, is exploited the idea to control and manipulate magnetism at the input/outputs of the functional blocks with an electric field at room temperature, wherein a fundamental unit of operation (for example, writing a state) takes energy less than or equal to 1 Atto-Joule. This corresponds to approximately 1.E-15 J/cm2 in terms of energy density. By this effect, some functional blocks may use spin, while others may use magnetic quantum pathway deflection. This is based on recent breakthroughs to enable room temperature, electric field control of magnetism [33, 38]. The basis for electric field control can be Bismuth Ferrite ($BiFeO_3$), and chemically substituted versions of this system, to tune its switching field thin films on a Si substrate that is exchange coupled to either a magnetic tunnel junction or a spin valve; the maximum switching voltage is in the range 0.5 V-1 V. Ferroelectric hysteresis of the spin valves may also be exploited for memory functions, and ferromagnetic resonance for amplification [39].

The quantum pathways carry the particles(with charges and spin) by ballistic conduction, so one must identify the phenomena by which the flows are enabled and controlled .Are exploited at the functional path gateways two properties of single-atom field-effect- transistors (FET) built by carbon nanotubes. First, as part of the parallel signalling system, they control the evacuation of the charge accumulated on porous island collectors and to offer the necessary potential switching capacity to control the "input charge flows" (referred to the charge accumulators). Next, these FET play the role of binding together specialized functional blocks. By building "pathways" between these functional blocks, the ballistic conduction offers the advantage of an extraordinary efficiency in the transport of quantum charge carriers. In this way, the heat developed as result of charge moving will be minimized and, in such a case, the control and signalling efficiency will be significantly increased.

For the memory functions, is exploited the dissymmetry created into the structured matter traversed by quantum pathways. This creates band gap zones where the charge can be stored; the granularity of such structures can potentially be brought down to single-electron charges. Thus, such structures can play the role of very dense memory functional blocks. The memories interconnect exploits the capacitance provided by porous charge accumulation islands that can be inter-connected by using 2D-graphene or carbon-nano-tubes (CNT). Therefore, by design, the charge collector zones could also implement spatial-temporal integration function. If the same porous accumulation island is targeted by opposite charge electron-hole carriers (triggered by controls or signalling commands), their coupling will generate excitons that cancel the charge [21]. An equivalent approach, but for the controlled emission of photons by quantum dots (so called optical "qubits") in semiconductors, require cavity quantum electrodynamics and a realization by Stranski-Krastanov strain induced formation [36].

Other phenomena which could be exploited to further improve the performance and energy use of the proposed architecture include:

-to improve insulation: the reciprocal annulations of opposite charge carrier particle coupling with generation of neutral excitons [40], and minimal conductivity materials;

- to better control the quantum pathways, the anomalous phase Berry effect (PBE) in ferromagnetic materials [22], that is the effect whereby when a current-carrying conductor is placed in a magnetic field, the Lorentz force "presses" its electrons against one side of the conductor [41];

-to enhance parallelization, the minimal high field degeneracy splitting;

-to increase memory capacity, use the quantum confinement effect [37];

-to enhance signalling flexibility: by shaping the atomic potential like Berry cones thanks to the effect known as phase Berry effect [22];

-create quantum entanglements such that cooling takes place when data is deleted [42].

All these phenomena rooted in advances in particle physics and materials research, allow us to affirm that the development of a new information processing architecture, based on condensed matter and quantum particles phenomena, will become a reality, and deliver both a high computation capacity and very high energy efficiency.

# V. Realization of the architectural components

This Section deals with the microelectronic and biology related design of the components of the proposed architecture.

The proposed architecture, as introduced in Section IV, combines two different types of synchronization:

i). Asynchronous processing, or "flow processing": it exploits the deflection of charge carriers pathways with accumulation of quantum charges at dedicated porous charge accumulation or confinement islands, as explained in Section V (see Figure 2);

ii). Synchronous processing: once the signalling has activated required functions and resources, the control path enables the programmability via state changes of automata and parameter selection, as explained in Section V (see Figure 3).

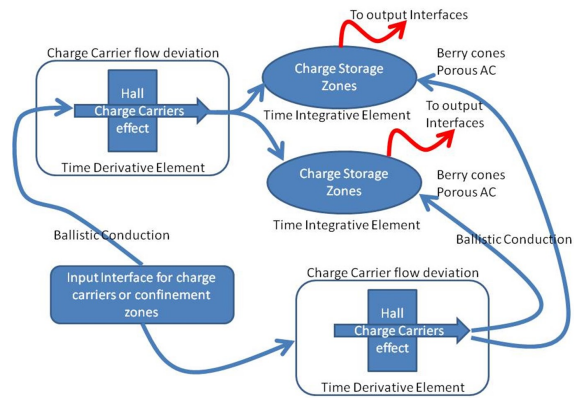

Figure 2: Quantum pathway path oriented asynchronous processing: interfaces and confinement zones; pathways based on ballistic conduction on 2D graphene; sensing and eventually threshold elements; charge carriers collector zones with or without sensing and eventually threshold elements. All these elements can be placed on a single or adjacent multiple graphene or other condensed matter layers

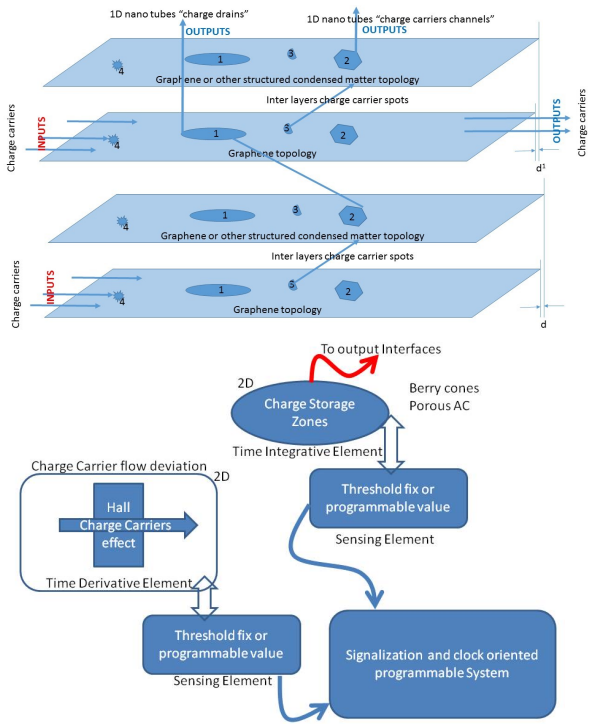

Figure 3: Quantum pathways synchronous processing: transition time based elements such as atomic FET's, and signalling by magnetic effects induced by electrical commands (see Section V); the clock circuits, signalling logic and programmable control signals reside in the conventional processor of Figure 1; integrated with the quantum based processing, they carry out the synchronous extraction of charge/ processed pathways.

The asynchronous processing performs at very high speed due to the propagation speed of quantum charge carriers inside the condensed matter structures. In these structures, the ferromagnetic deflection elements may belong, or not, to the signalling system. When the deflection elements do not belong to the signalling, they can be combined with fixed threshold functions, into deflection lattices, and even with sensing / actuators functions.

Customization is possible. This may happen obviously first via the layout of interconnections between the structural elements around the functional blocks (Figure 4); these elements include the confinement and collector areas for quantum charge carriers (including the special case of input and output gateways to the processing system), quantum pathways dedicated to propagation, ferroelectric deflection elements, and atomic FET's. Next, the functional blocks and interconnects can be stacked to couple or decouple the quantum interactions; obvious cases include the stacking of graphene, ferromagnetic or CNT sheets .And finally, customization happens by the nature and properties of the functional blocks, which are orchestrated by the signalling process.

Figure 4: Interconnect structure: Each layer can be represented as a "informational compact": 1 represents charge collector islands; 2 are sub-lattice able to deflect the charge carriers flows, 3 and 4 are induced structural defects or QHE based elements, and d, d1 are the Moiré interference steps between different layers.

Scaling offers interesting new research opportunities. Initial realizations can rely on graphene, CNT or ferroelectric layers combined with porous quantum charge accumulation zones, as explained above. A path towards downscaling by one or two orders of magnitude exists by aiming at collecting single quantum charges or electrons in atomic local density of states (LDOS) band gaps till sufficient charges are collected due to the quantum pathways feeding and transforms. This could happen as the dissymmetry build-up created on mono-atomic layers creates band gap zones where the charge carriers can be collected. In this last case, it is yet unknown however if second order conduction processes exist, affecting latency of the quantum wells. Needless to say, the described scalability also affects the reduction of the energy consumption

Integrated sensing and sensor information pre-processing are offered as well, thanks mostly to the functional blocks replicating biological sensing processes, many of which have embedded closed loop processing. This is enabled because along quantum pathways, one can collect both the instantaneous quantum charge levels and spin, but also their integral values measured at charge collectors' level. The tuning is by the coupling of the sensing elements by small electric potentials or magnetic fields (determined by the electrical-magnetic interaction effect at controller level). Geim and Novoselov [18] have already demonstrated these strange close-loop quantum phenomena to take place in the very close neighborhood of a quantum charge accumulation island (at a distance of max. 10 successive 2D layers).

Finally, the proposed architecture offers a high degree of redundancy, with signalling paths serving the reconfiguration, thanks to the controlled quantum pathway's splits and mergers.

# VI. Comparison of the programmability in current ISA processors and quantum biological systems

In order to better clarify the analogies and differences that exists between classical computer programmability features, reflected by the Instruction Set Architecture (ISA), and the proposed quantum biological processing system, a comparison table is provided (Table 1).

State of the art programmable computers use three main types of instructions: executable, decision and repetitive. On top of that, structured programming technologies have been developed, such as object oriented programming, which offer a more generic way to implement algorithms and to increase programming productivity. The main constructs in the programming languages are: data structures, routines, functions, and objects.

Even if the control processor in Figure 1 has an ISA architecture and the above constructs, an issue is how quantum biological information systems can be modeled and implement similar techniques? For the formalization and modeling of the new systems, we suggest the use of generalized colored algebras combined with geometry and graph theory [43-45]. First, geometry and physical interaction models specify the physical structures, just as in current microelectronic CAD systems, except that electron charge equations must be supplemented by quantum charge [46] and spin [47] equations, and that the substrate properties must be supplemented accordingly. Next, colored algebras [45,48], with its labeled initial symbols, groups, rings, and their various properties (especially commutativity or non-commutativity, and transitivity or non-transitivity ) and operands, can formalize the interactions inside functional blocks while classifying them by attribute ranges linked to energy levels. Already now exist compilers for simpler versions, called colored Petri nets [44], which may serve as a starting point to model and simulate the functional blocks. Finally, for simulation, quantum simulators already exist capturing the behaviors of quantum systems, such as the MOQUASIMS (Memory enabled optical quantum simulator) [34.]

Table 1: A comparison table between instruction set architecture computer's elements and their equivalence in the proposed quantum biological computer architecture.

| Elements in ISA Computers based on Turing machine architecture- | Equivalent elements in information processors based on quantum biological architecture" |
|---|---|
| Executable instructions | Realization by quantum charge collector islands, propagation "pathways" where conduction is performed in a ballistic mode, deflection elements (lattice, Berry cones, FET's, and Moiré based sub-lattice implementing spatial interferential computing). The executability results from the |

| | | | |
|---|---|---|---|
| | signalling system reserving the required resources above, required to execute an instruction. The 2-D or 3-D stacking of such elements (see Figure 4) allows to implement complex instructions. | | *differential signals* or to compute derivatives (QHE is dependent of speed of charge carriers, but controlled by the local magnetic field).<br><br>3) By designing a close loop pathway inside a functional block, a *time dependent* threshold can be implemented. |
| Decision instructions | 1) Use of different techniques for sensing *integral* signals, such as the cumulated charge collected at porous island ( or Berry cones), coupled with decision thresholds set by the control information path (Figure 1).<br><br>These thresholds can have:<br>• "fixed" values, as result of the structural design, or depend strictly upon the dimensions of atoms involved which selected via a Moiré sub-lattice constant, or:<br>• "variable" values, when inside the atomic structure some sensing elements are controlled by voltage or magnetic field.<br><br>2) As QHE is dependent on the speed of the quantum charge carriers, the QHE effect in particular will permit to sense | Repetitive instructions | Homonymic and toponymic structures for the quantum pathways implement repetitive instructions. A Spin-FET or a sensing element provide access/exit control of a carrier flow inside a reverberant functional block (see Figure 5) .Such structures function like "mirrors" for the traveling charge carriers and, based on thresholds inserted on the pathways, these carriers will be ejected from the reverberant system . |
| | | Data structures | Data structures are normally defined in the control processor (Figure 1) and adapted there to their injection via the electric-magnetic interaction (Section IV). If not, semantic allocation attributes can be put in equivalence to cortex reflection zones in a dedicated biology inspired functional block. |

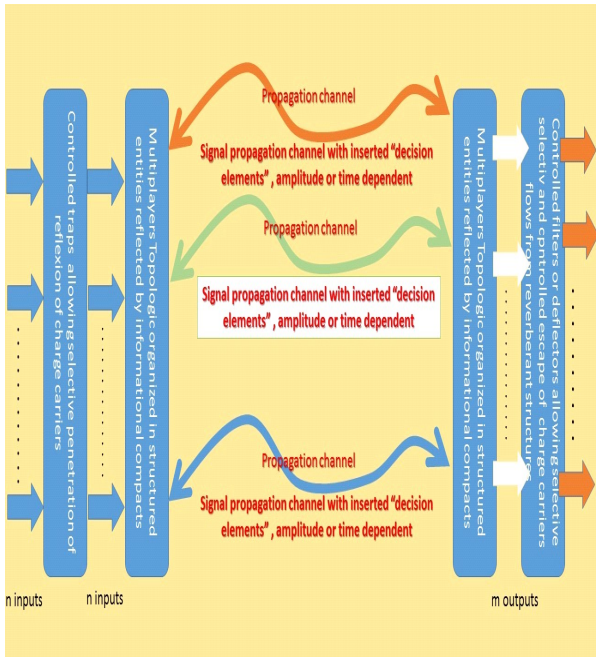

Figure 5: Schematic representation of reverberant structures. The propagation channels represent selective quantum pathways for charge carriers, not represented into the picture.

## VII. Open research issues

The paper tries to offer a visionary perspective on the evolution of information processing systems, first by combining conventional Turing machines with a new quantum biological architecture, and next by unifying the normally disjoint views of information and energy via the use of quantum processing. Thinking about the time perspective for an implementation of this vision, it can be observed that much of the underpinning research in particle physics, biology and material science has recently accelerated, creating a moderate optimism on realizations in a foreseeable future.

But the remaining research issues are many; some of the most important include the:

-identification of additional biological processes, which have been investigated well enough experimentally, but the information processing capabilities of which have not yet been recognized; the field of pattern recognition offers many cues, but knowledge is insufficient on biological and cognitive processes involved in many other tasks;

- -necessity to improve the nanotechnology processes related to designing and stacking graphene, CNT and ferroelectric layers with localized quantum charge accumulation islands, despite laboratory field trials
- -stacking of atomic layers;
  -integration of ferromagnetic gateways on graphene or CNT;
- -analysis of the "life time" of some of the sub-particles resulting from the interactions, and elucidation of a model based on confinement and junction of such particles, that in fact becomes observable for a certain period of time;
- - reciprocal annihilation of charge carriers, resulting in excitons or Majorana sub-particles; this phenomenon is important to study in order to find determine the energetic model for the residual mass and mass less particles;
- -the mathematical formalization, synthesis and simulation for building the functional blocks and other processing elements; however, tools exist already now, and many biological processes have been studied well enough experimentally to offer a significant modeling base.

## Conclusions

The proposed architecture is certainly visionary and its realism hinges on the extent to which the limitless diversity of information processing tasks is not overwhelming the combinatorics and programmability of functions inspired from biology and embedded into this design. In many ways, the history of computers sofar has faced the same challenge, but without exploiting neither the richness of biology or the processing speed offered by quantum particles.

In terms of energy savings and clock speed gains, they are both huge and much higher than in state-of-the art silicon structures. This is the consequence of the very low electric resistance along quantum pathways, and, also, of the very high thermal conductivity graphene / CNT layers.

Moreover, by building up a cellular based architecture, with parallel-linked processing elements, it is very similar to the case of living systems that are composed by aggregated live cell structures.

Specificity, computation speed and low energy consumption become simpler to be conceived. Unfortunately, as result of the diversity of functional blocks, and of possible limits on possibilities to assemble them, the design complexity increases in comparison with classical computation systems. Fortunately, the modeling of the new proposed architecture benefit from the use of hitherto neglected algebraic formalisms.

# Acknowledgments

The authors thank for the prolific discussions and debates that they had in the frame of the EU funded MP1004 COST project, and with several advanced research programs or visionaries in industry worldwide.